\documentclass[twocolumn]{article}
\let\ifapj\iffalse
\let\ifarxiv\iftrue
\let\iflocal\iffalse
\usepackage{ifxetex,ifluatex}
\ifxetex\else\ifluatex\else\usepackage[utf8]{inputenc}\fi\fi
\usepackage{scrlfile}
\ifarxiv\usepackage{font-termes}\fi
\iflocal\usepackage{font-termes}\fi
\usepackage{etoolbox,ifthen}
\ifapj
  \usepackage[strict]{patch-apj}
  \usepackage{acro,appendix,csquotes,hyperref,siunitx}
\fi
\ifboolexpr{bool{arxiv} or bool{local}}{
  \usepackage{siunitx}
  \usepackage{basic,physics,astro}
  \usepackage[graytext]{draft}
  \usepackage[mode=biblatex, hyperref]{apjbib}
  \ExecuteBibliographyOptions{embedlinks}
  \addbibresource{tde.bib}
}{}
\ifluatex\hypersetup{pdfencoding=auto}\fi
\makeatletter
\ifxetex\else\ifluatex\else
  \@ifpackageloaded{astro}{\let\ast@symit\textit}{}
\fi\fi
\makeatother
\DeclareAcronym{TDE} {short=TDE,  long=tidal disruption event}
\DeclareAcronym{AGN} {short=AGN,  long=active galactic nucleus,                short-indefinite=an,                long-indefinite=an, long-plural-form=active galactic nuclei}
\DeclareAcronym{MHD} {short=MHD,  long=magnetohydrodynamic,                    short-indefinite=an, short-plural=,                     long-plural=s}
\DeclareAcronym{RIAF}{short=RIAF, long=radiatively inefficient accretion flow}
\newcommand*\Mh{M_\su h}
\newcommand*\Ms{M_\star}
\newcommand*\rs{r_\star}
\newcommand*\rt{r_\su t}
\newcommand*\rp{r_\su p}
\newcommand*\rg{r_\su g}
\newcommand*\La{L_\su a}
\newcommand*\Lc{L_\su c}
\newcommand*\LE{L_\su E}
\newcommand*\smc{\dot M_\su s}
\newcommand*\dmc{\dot M_\su d}
\newcommand*\amc{\dot M_\su a}
\newcommand*\emc{\dot M_\su E}
\newcommand*\mcr{\smc/\dmc}
\newcommand*\tmb{t_\su{mb}}
\newcommand*\torb{t_\su{orb}}
\newcommand*\tinfl{t_\su{infl}}
\newcommand*\tcool{t_\su{cool}}
\newcommand*\tdec{t_\su{dec}}
\newcommand*\tdecfid{10}
\ifapj\let\edit\relax\else\usepackage{xcolor}\fi
\expandafter\def\csname editcolor1\endcsname{blue!60!cyan}
\newcommand*\edit[2]{\textcolor{\csname editcolor#1\endcsname}{#2}}
\begin{document}

\title{Light curves of tidal disruption events in active galactic nuclei}

\ifapj
  \author[0000-0001-5949-6109]{Chi-Ho Chan}
  \affiliation{Racah Institute of Physics, Hebrew University of Jerusalem,
  Jerusalem 91904, Israel}
  \affiliation{School of Physics and Astronomy, Tel Aviv University,
  Tel Aviv 69978, Israel}
  \author[0000-0002-7964-5420]{Tsvi Piran}
  \affiliation{Racah Institute of Physics, Hebrew University of Jerusalem,
  Jerusalem 91904, Israel}
  \author[0000-0002-2995-7717]{Julian~H. Krolik}
  \affiliation{Department of Physics and Astronomy, Johns Hopkins University,
  Baltimore, MD 21218, USA}
\fi

\ifboolexpr{bool{arxiv} or bool{local}}{
  \author[1,2]{Chi-Ho Chan}
  \author[1]{Tsvi Piran}
  \author[3]{Julian~H. Krolik}
  \affil[1]{Racah Institute of Physics, Hebrew University of Jerusalem,
  Jerusalem 91904, Israel}
  \affil[2]{School of Physics and Astronomy, Tel Aviv University,
  Tel Aviv 69978, Israel}
  \affil[3]{Department of Physics and Astronomy, Johns Hopkins University,
  Baltimore, MD 21218, USA}
}{}

\date{October 30, 2020}

\shorttitle{Light curves of TDEs in AGNs}
\shortauthors{Chan et al.}
\pdftitle{Light curves of tidal disruption events in active galactic nuclei}
\pdfauthors{Chi-Ho Chan, Tsvi Piran, Julian Krolik}

\begin{abstract}
The black hole of \iacl{AGN} is encircled by an accretion disk. The surface
density of the disk is always too low to affect the tidal disruption of a star,
but it can be high enough that a vigorous interaction results when the debris
stream returns to pericenter and punches through the disk. Shocks excited in
the disk dissipate the kinetic energy of the disk interior to the impact point
and expedite inflow toward the black hole. Radiatively efficient disks with
luminosity \num{\gtrsim e-3} Eddington have a high enough surface density that
the initial stream--disk interaction leads to energy dissipation at a
super-Eddington rate. Because of the rapid inflow, only part of this dissipated
energy emerges as radiation, while the rest is advected into the black hole.
Dissipation, inflow, and cooling balance to keep the bolometric luminosity at
an Eddington-level plateau whose duration is tens of days, with an almost
linear dependence on stellar mass. After the plateau, the luminosity decreases
in proportion to the disk surface density, with a power-law index between $-3$
and $-2$ at earlier times, and possibly a steeper index at later times.
\end{abstract}
\acresetall

\section{Introduction}

When a star of mass $\Ms$ and radius $\rs$ approaches a black hole of mass
$\Mh$ on an orbit whose pericenter $\rp$ is comparable to or smaller than the
tidal radius $\rt\eqdef\rs(\Ms/\Mh)^{-1/3}$ \citep{1975Natur.254..295H}, the
tidal gravity of the black hole rips the star asunder
\citep[e.g.,][]{1988Natur.333..523R}, leading to \iac{TDE}. The disruption
takes place very close to the black hole: in units of the gravitational radius
$\rg\eqdef G\Mh/c^2$, the tidal radius is
\begin{equation}
\rt\approx50\,\rg\,
  \biggl(\frac\Mh{\SI{e6}{\solarmass}}\biggr)^{-2/3}
  \biggl(\frac\Ms{\si{\solarmass}}\biggr)^{-1/3}
  \biggl(\frac\rs{\si{\solarradius}}\biggr).
\end{equation}
The mass return time, determined by the specific energy of the most bound
debris, is the timescale on which this part of the debris returns to
pericenter:
\begin{equation}
\tmb\sim\SI{40}{\day}\times
  \biggl(\frac\Mh{\SI{e6}{\solarmass}}\biggr)^{1/2}
  \biggl(\frac\Ms{\si{\solarmass}}\biggr)^{-1}
  \biggl(\frac\rs{\si{\solarradius}}\biggr)^{3/2}
  \biggl(\frac\rp\rt\biggr)^3.
\end{equation}
General relativistic simulations of \acp{TDE} starting from stars with
realistic structures reveal that these quantities are accurate only within a
factor of \num{\sim2} \citep{2020arXiv200103501R}.

The black hole of \iac{AGN} is girdled by an accretion disk. When \iac{TDE}
happens in the vicinity of such a black hole, the passage of the star through
the disk leaves the disk largely intact, and the disruption proceeds as if the
disk were absent. However, upon return to the vicinity of the black hole, the
debris of the disrupted star, being much more extended and dilute than the
star, can interact with the disk in a more interesting way
\citetext{\citealp{1994ApJ...422..508K}\multicitedelim
\bibstring{seealso}\space\citealp{2017MNRAS.469..314K}}.

As we shall argue later, the stream typically has much more inertia than the
disk, so it acts as an immovable obstacle to disk rotation. Shocks form where
the disk runs into the stream, and the shocks dissipate disk kinetic energy.
When the disk surface density is large enough, the dissipation rate can be
highly super-Eddington. This sort of shock dissipation could serve as the
energy source for a bright flare different from the energy sources in ordinary
\acp{TDE}, namely, accretion of any rapidly circularized debris
\citep[e.g.,][]{1988Natur.333..523R} and shocks due to stream self-interaction
\citep{2015ApJ...806..164P}.

In \citet{2019ApJ...881..113C}, we performed the first hydrodynamics
simulations of the collision between the debris stream of \iac{TDE} and the
pre-existing accretion disk of \iac{AGN}. We found that, as long as the stream
is much heavier than the disk, our simulation results are sensitive only to how
dense the disk is, not how dense the stream is. This observation allows us to
extrapolate our simulation results to even heavier streams and make
quantitative predictions, even though our simulations did not explicitly cover
that regime.

We begin by highlighting the most salient simulation results in
\cref{sec:summary}. Based on these results, we estimate in
\cref{sec:luminosity} the bolometric light curve expected when a debris stream
tears through the disk. In \cref{sec:light curves}, we scrutinize the
dependence of the light curve on black hole mass and disk properties, and we
present sample light curves. This discussion is followed by a comparison with
previous theoretical models and observations in \cref{sec:discussion}. We end
with our conclusions in \cref{sec:conclusions}.

\section{Summary of simulations}
\label{sec:summary}

\begin{figure}
\includegraphics{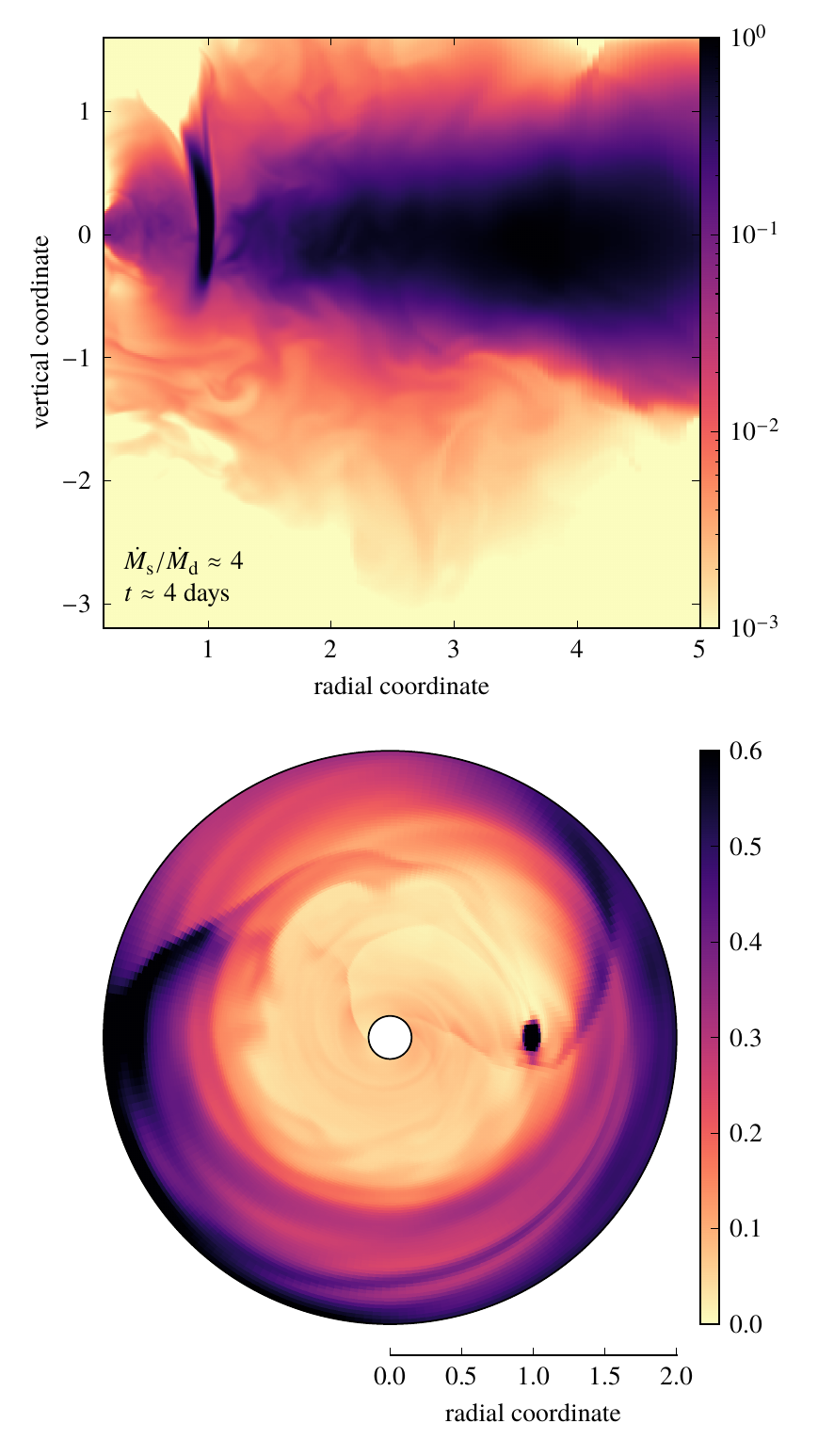}
\caption{Density snapshot of one simulation in \citet{2019ApJ...881..113C}. The
top panel is a poloidal slice through the stream impact point with colors in
logarithmic scale. The pericenter part of the stream, visible as a vertical
structure, penetrates the disk with little difficulty, and energy dissipation
makes the disk around the stream geometrically thick. The bottom panel is a
midplane slice with colors in linear scale. The dark dot is a section of the
stream, and spiral shocks are discernible interior to the stream. The density
unit is arbitrary.}
\label{fig:density}
\end{figure}

In \citet{2019ApJ...881..113C}, we simulated the collision between the debris
stream of \iac{TDE} and the pre-existing accretion disk of \iac{AGN}. As
illustrated in \cref{fig:density}, a parabolic stream, representing the
returning debris, strikes the disk perpendicularly from above as it reaches
pericenter. To study late-time behavior, here we extend the simulations from
the original duration of \SI{\sim12}{\day} to double that. The longer duration
is still a fraction of $\tmb$, so it is reasonable to keep the stream
conditions constant.

We adopt as fiducial parameters $\Ms=\si{\solarmass}$, $\rs=\si{\solarradius}$,
and $\rp=\rt$. Since our interest is in disk physics, the relevant timescale is
the disk orbital time at the impact point, which is by definition $2\pi$ times
the stellar dynamical time:
\begin{equation}\label{eq:orbital time}
\torb
  =2\pi\biggl(\frac{\rp^3}{G\Mh}\biggr)^{1/2}
  =2\pi\biggl(\frac{\rs^3}{G\Ms}\biggr)^{1/2}\approx\SI{3}{\hour}.
\end{equation}
The independence of $\torb$ on $\Mh$ means that our results, quoted in units of
days, are valid for all $\Mh$.

The most important parameter governing the collision is $\mcr$, the ratio of
the mass current of the stream to the mass current of the disk rotating under
the stream footprint. We shall see in \cref{sec:light curves} that
$\mcr\gtrsim1$ typically; in other words, the stream has too much inertia to be
affected by the disk, and it obstructs the rotating disk instead.

\begin{figure*}
\includegraphics{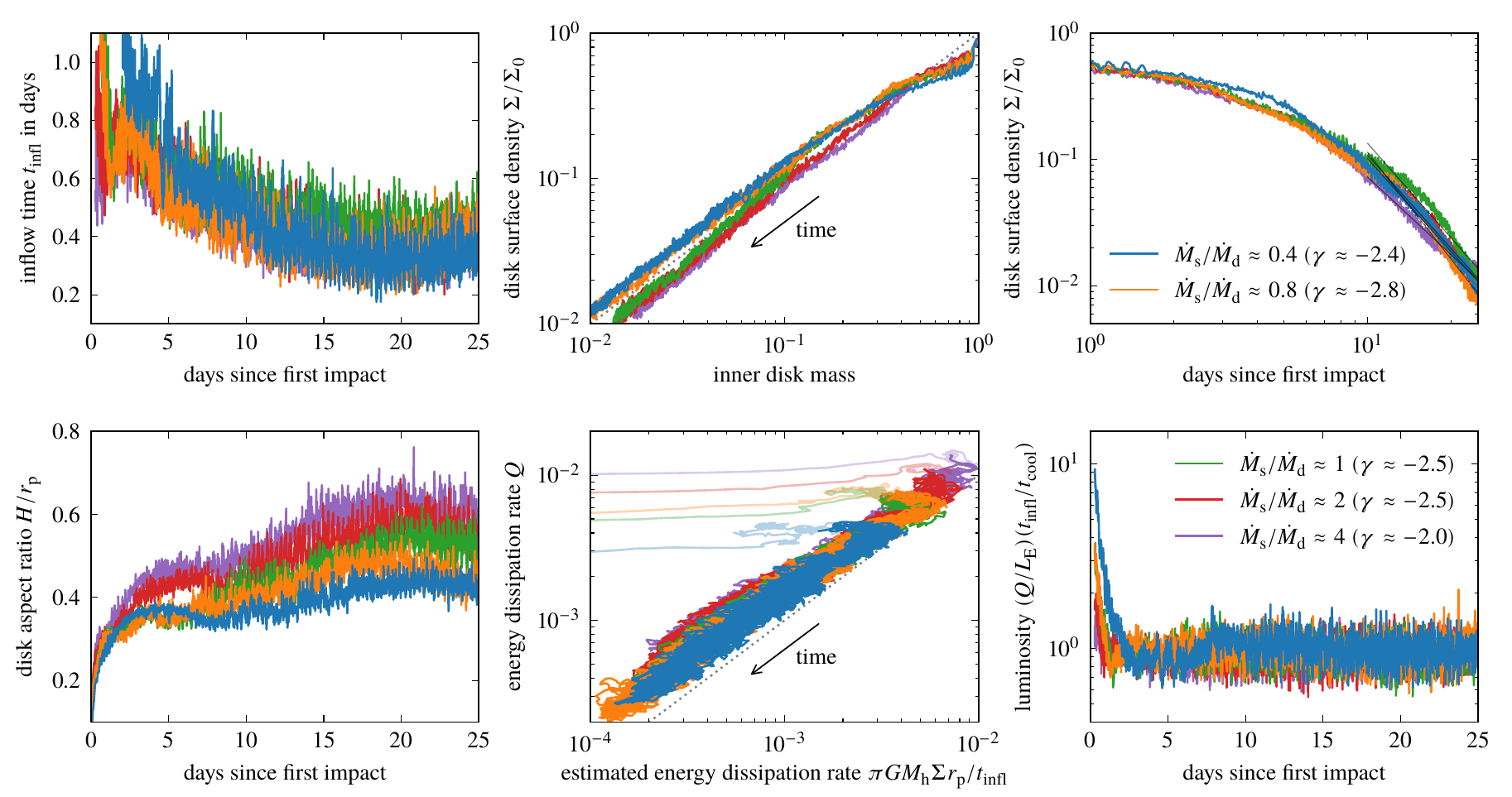}
\caption{Diagnostics of the simulations in \citet{2019ApJ...881..113C}, doubled
in duration to study late-time behavior. The legend in the right column applies
to all panels. The top-left panel shows the inflow time, that is, the inner
disk mass divided by the inflow rate onto the black hole. The top-center panel
shows the correlation between the inner disk mass and the disk surface density
$\Sigma$ at the impact point, measured from the midplane to infinity; both
quantities are normalized to their initial values, and the dotted line marks a
one-to-one ratio. The top-right panel shows $\Sigma$ as a fraction of its
initial value; the thin lines are power-law fits to late-time behavior and
$\gamma$ in the legend is the power-law index. The bottom-left panel shows the
disk aspect ratio at the impact point. The bottom-center panel compares the
energy dissipation rate estimated with \cref{eq:energy dissipation rate} and
that measured directly from the simulations, both in arbitrary units;
early-time data are plotted in a lighter color to highlight the late-time
trend, and the dotted line marks a one-to-one ratio. The bottom-right panel
shows the estimate of bolometric luminosity as defined in
\citet{2019ApJ...881..113C}, divided by the Eddington luminosity. The more
accurate \cref{eq:luminosity} reduces to this estimate in the limit
$\tinfl/\tcool\lesssim1$, which likely holds at early times when $\Sigma$ is
still relatively large.}
\label{fig:simulation}
\end{figure*}

The collision excites multiple spiral shocks in the disk inside the stream
impact radius. The shocks remove angular momentum and dissipate kinetic energy,
thereby clearing out the inner disk within several $\torb$. \Ac{MHD} stresses
are ignored in the simulations, but they act much more slowly than shocks. The
top-left panel of \cref{fig:simulation} tells us that the inflow time $\tinfl$,
or the time it takes shocks to redirect gas toward the black hole, gradually
declines over the course of the event, from \SI{\sim1}{\day} at the start to
\SI{\sim0.4}{\day} at \SI{\gtrsim10}{\day}.

As impact with the disk shaves off part of the stream, the inner disk mass is
partially replenished with stream material. The inner disk mass is determined
by the balance between inflow and replenishment. The longer simulations make
clear this dynamic. The top-center panel of \cref{fig:simulation} demonstrates
that the inner disk mass drops roughly in proportion to the disk surface
density $\Sigma$ at the impact point. The top-right panel shows that
$\gamma\eqdef\ods{\ln\Sigma}{\ln t}$ lies between $-3$ and $-2$ toward the end
of the simulations, and the decrease of $\Sigma$ gradually steepens over time.
We emphasize that the time-evolution of $\Sigma$ is due to mechanisms wholly
unrelated to those determining the power-law decay of the mass return rate at
$t\gtrsim\tmb$.

The dissipated energy heats up the inner disk. As portrayed in the bottom-left
panel of \cref{fig:simulation}, heating raises the disk aspect ratio at the
impact point to $H/\rp\sim\tfrac12$. The cooling time $\tcool$ is the time the
inner disk takes to release its internal energy as radiation, including the
time needed to convert internal energy to radiation, and the time needed for
radiation to escape through free streaming, diffusion, or vertical advection
due to convection and magnetic buoyancy. Because electron scattering dominates
the opacity in these circumstances, the diffusion time is $\mathrelp\sim\tau
H/c$, where $\tau=\Sigma\kappaT$ is the Thomson optical depth and $\kappaT$ is
the cross section per mass for Thomson scattering. The diffusion time is long
because $\tau$ drops rather slowly while $H/\rp\lesssim1$ is much larger than
in the unperturbed disk; in fact, the diffusion time is typically
$\mathrelp\gtrsim\tinfl$ at early times. The cooling time could be even longer
because it also takes into account the radiation time. This means most of the
radiation is trapped in the inflow toward the black hole and only a small
fraction escapes.

\section{Estimation of bolometric luminosity}
\label{sec:luminosity}

The swift inflow in the perturbed inner disk could produce a flare. Similar to
a steady-state disk, the energy of the flare ultimately comes from the release
of the gravitational energy of the disk gas. The contrast with a steady-state
disk is twofold: first, dissipation and inflow happen on much shorter
timescales; second, mass resupply from the stream allows more energy to be
released than the gravitational energy possessed by the unperturbed disk.

We proposed in \citet{2019ApJ...881..113C} a crude estimate for the bolometric
luminosity $\Lc(t)$ of the collision-induced flare. Here we use the modified
form
\begin{equation}\label{eq:luminosity}
\Lc\sim Q\min(1,\tinfl/\tcool),
\end{equation}
where $Q(t)$ is the energy dissipation rate. This equation encapsulates a
competition between the inner disk releasing its internal energy as radiation,
represented by $\tcool$, and the inflow sweeping this radiation into the black
hole, represented by $\tinfl$. When $\tinfl/\tcool\lesssim1$, the ratio
$\tinfl/\tcool$ estimates the fraction of dissipated energy the inner disk
manages to radiate away; when $\tinfl/\tcool\gtrsim1$, all the dissipated
energy is radiated away before the gas is accreted, so $\Lc\sim Q$.
\Cref{eq:luminosity} is only an estimate; the true luminosity must be
determined by performing time-dependent radiative transfer on the actual
distribution of heating and opacity.

The top-left panel of \cref{fig:simulation} and \cref{eq:orbital time} together
tell us that
\begin{equation}\label{eq:inflow time}
\tinfl\sim4\torb.
\end{equation}
Gas does not plunge into the black hole; rather, it spirals inward along
trajectories that shift from quasi-circular near the impact radius to more
eccentric at smaller radii. When disk material strikes the obstacle posed by
the stream, it loses a large part of both its angular momentum and its energy;
when it is deflected by shocks at smaller radii, the fractional loss of angular
momentum is greater. For this reason, the heating rate as a function of time is
reasonably estimated by the orbital kinetic energy of gas orbiting at the
impact radius per inflow time:
\begin{equation}\label{eq:energy dissipation rate}
Q\sim\frac{\pi G\Mh\Sigma\rp}\tinfl\sim
  \frac\tau{8\pi}\biggl(\frac\rp\rg\biggr)^{-1/2}\frac\torb\tinfl\LE,
\end{equation}
where $\LE=4\pi G\Mh c/\kappaT$ is the Eddington luminosity. The bottom-center
panel of \cref{fig:simulation} demonstrates that, apart from an initial
adjustment phase, this estimate is accurate to within a factor of \num{\sim2}.

In estimating $\tcool$, we ignore the contributions of radiation time and
vertical advection for simplicity. In fact, finite radiation time slows cooling
while vertical advection speeds it, hence the two effects partially cancel. The
inner disk starts out optically thick, so
\begin{equation}\label{eq:cooling time}
\tcool\sim\frac{\tau H}c
  \sim\frac\tau{4\pi}\biggl(\frac\rp\rg\biggr)^{-1/2}\torb,
\end{equation}
where we used $H/\rp\sim\tfrac12$ in the second step. The inner disk eventually
becomes so depleted that $\tau\lesssim1$ and radiation can escape freely, at
which point $\Lc\sim Q$. This free-streaming limit is reached only when
$\tinfl/\tcool\gtrsim1$, since $\tinfl\gtrsim\torb$ always and
$\tcool\lesssim\torb$ in this limit. Considering that $\Lc$ is capped by the
minimum function in \cref{eq:luminosity} long before the free-streaming limit
kicks in, the limit is irrelevant and we can use \cref{eq:cooling time} for all
$\tau$.

We see from \cref{eq:cooling time} that, as long as the inner disk maintains a
high enough $\Sigma$ to make
\begin{equation}\label{eq:plateau condition}
\tau\gtrsim4\pi\biggl(\frac\rp\rg\biggr)^{1/2}\frac\tinfl\torb,
\end{equation}
we have $\tinfl/\tcool\lesssim1$. Combining \cref{eq:luminosity,eq:energy
dissipation rate,eq:cooling time} in this regime yields
\begin{equation}\label{eq:plateau}
\Lc/\LE\sim\tfrac12.
\end{equation}
This is our main result: The bolometric light curve of \iac{TDE} in \iac{AGN}
starts off with an Eddington-level plateau. Our simulations confirm the
validity of \cref{eq:plateau}: $Q\tinfl/\tcool$ in the bottom-right panel of
\cref{fig:simulation} quickly stabilizes to $\mathrelp\sim\LE$. The remarkable
constancy of $\Lc$ is due to the cancellation of $\tinfl$ and $\tau$ in the
derivation of \cref{eq:plateau} \citep[see also][]{2010ApJ...709..774K}. When
the inner disk is cleared out to the point that \cref{eq:plateau condition} is
violated and $\tinfl/\tcool\gtrsim1$, the luminosity falls as $\Lc\sim
Q\propto\Sigma$. Both the plateau duration and the post-plateau $\Lc$ depend on
the time evolution of a single parameter, $\Sigma$.

Because the stream feeds the inner disk, the plateau duration is not limited by
the initial inner disk mass and can be $\mathrelp\gg\tinfl$; the plateau
duration will be estimated in \cref{sec:light curves}. A plateau may not appear
if radiation time, convection, or magnetic buoyancy modifies the functional
form of $\tcool$.

Our calculations give us only a crude estimate of the bolometric luminosity in
some time-averaged sense. Accurate predictions of the light curve and the
spectrum should be based on detailed radiation \acp{MHD} simulations.

\section{Light curves}
\label{sec:light curves}

We consider two disk models for the unperturbed disk, depending on how the
unperturbed accretion rate $\amc\eqdef\La/(\eta c^2)$ compares with the
Eddington accretion rate $\emc\eqdef\LE/(\eta c^2)$, where $\eta=0.1$ is the
fiducial radiative efficiency of a radiatively efficient disk. Any time-steady
disk satisfies $\amc\sim 4\pi\rp\Sigma_0v_R$, where $\Sigma_0$ and $v_R$ are
the unperturbed surface density and radial velocity respectively. The value of
$\Sigma_0$ is fixed by $\dmc$; therefore, a choice of disk model boils down to
a choice of $v_R$. The inflow mechanism in the unperturbed disk is the outward
transport of angular momentum by internal stresses, but once the stream strikes
the disk, inflow is driven by spiral shocks instead.

If $\amc/\emc\gtrsim\num{e-3}$, we assume the disk is geometrically thin,
optically thick, and radiatively efficient \citep{1973A&A....24..337S} with
fiducial stress parameter $\alpha=0.1$. Both the disk aspect ratio and
$v_R/v_\phi$ at the impact point are \num{\ll1} in this model, $v_\phi$ being
the orbital velocity. As a result, $\Sigma_0$ is large, and the initial energy
dissipation rate $Q_0$, obtained by substituting the unperturbed surface
density $\Sigma_0$ into \cref{eq:energy dissipation rate}, is large as well.

If $\amc/\emc$ is any lower, the disk could be a geometrically thick, optically
thin, \ac{RIAF} \citep[e.g.,][]{1977ApJ...214..840I, 1982Natur.295...17R,
1994ApJ...428L..13N}. Parameterizing the radial velocity in a geometrically
thick disk as $v_R\sim\alpha'v_\phi$, we can write
\begin{equation}
\kappaT\Sigma_0\sim\frac1{\alpha'\eta'}\frac{\La/\LE}{(\rp/\rg)^{1/2}}.
\end{equation}
We emphasize that $\alpha'$ is merely a parameterization; for simplicity we
take $\alpha'=0.1$. \Citet{2017ApJ...844L..24R} performed general relativistic
\acp{MHD} simulations of \acp{RIAF} assuming that electrons are heated by
Coulomb scattering off ions at a rate derived empirically from solar wind
measurements \citep{2015MNRAS.454.1848R}. They found that the effective
radiative efficiency is $\eta'\sim\num{e-2}$ for $\amc/\emc\sim\num{e-5}$,
suggesting that accretion may proceed at efficiencies close to radiatively
efficient values even at very low accretion rates. We adopt this value of
$\eta'$ below. The much larger $v_R/v_\phi$ in \iac{RIAF} compared to a
radiatively efficient disk means that $\Sigma_0$ and hence $Q_0$ are much
smaller.

\begin{figure}[p]
\includegraphics{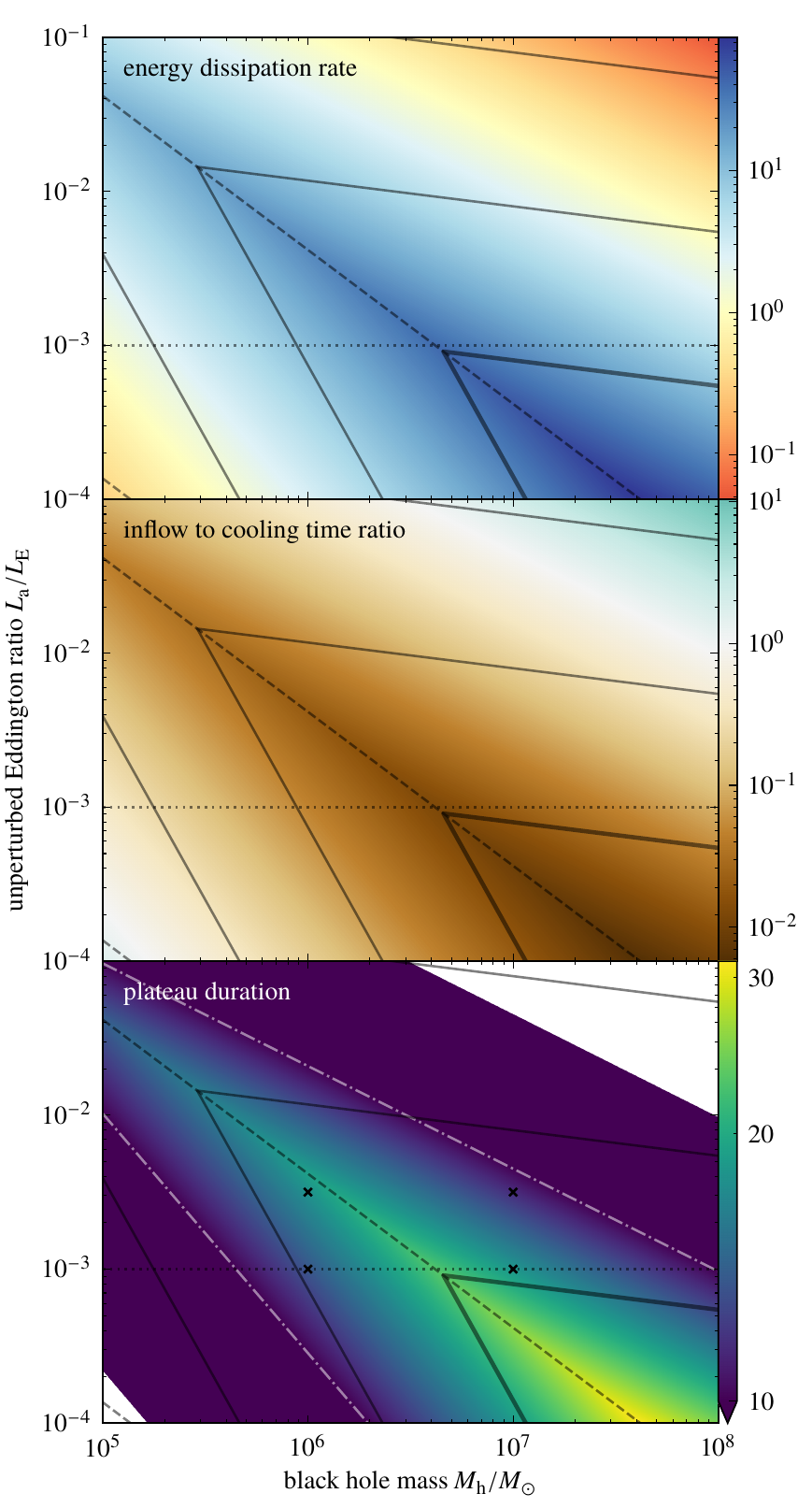}
\caption{\ifapj\scriptsize\fi
Color contours of three properties of the perturbed inner disk, as functions of
the black hole mass $\Mh$ and the Eddington ratio $\La/\LE$ of the unperturbed
disk, with other parameters fixed at their fiducial values
(\cref{sec:summary,sec:light curves}). This figure portrays the case in which
the unperturbed inner disk is a radiatively efficient disk, while
\cref{fig:riaf} illustrates the case of \iac{RIAF}. The top panel shows the
initial energy dissipation rate $Q_0$ divided by the Eddington luminosity
$\LE$. The middle panel shows the initial value of the ratio of inflow to
cooling time $\tinfl/\tcool$. The bottom panel shows the plateau duration in
units of days (\cref{sec:plateau duration}). The blue regions outside the
dash--dotted lines have only an upper limit, while the empty regions have no
plateau at all because $\tinfl/\tcool\gtrsim1$ initially. The crosses are the
values of $(\Mh,\La/\LE)$ for which light curves are shown in \cref{fig:light
curve}. In all panels, the lower dashed line at
$\La\sim\SI{1.7e39}{\erg\per\second}$ separates two opacity regimes of the
unperturbed disk: opacity is dominated by free--free absorption below and
electron scattering above. The upper dashed line at
$\La\sim\SI{5e41}{\erg\per\second}$ separates two pressure regimes of the
unperturbed disk: pressure is dominated by gas below and radiation above. The
thick gray chevron is the contour for $\mcr=1$, and the two thin ones above it
are for $\mcr=10$ and $\mcr=100$ respectively. Our approximation that the
stream has greater inertia than the disk holds for $\mcr$ above a few tenths
\citep{2019ApJ...881..113C}. The horizontal dotted line marks the value of
$\La/\LE$ below which the radiatively efficient disk may become \iac{RIAF}.}
\label{fig:ss73}
\end{figure}

\begin{figure}
\includegraphics{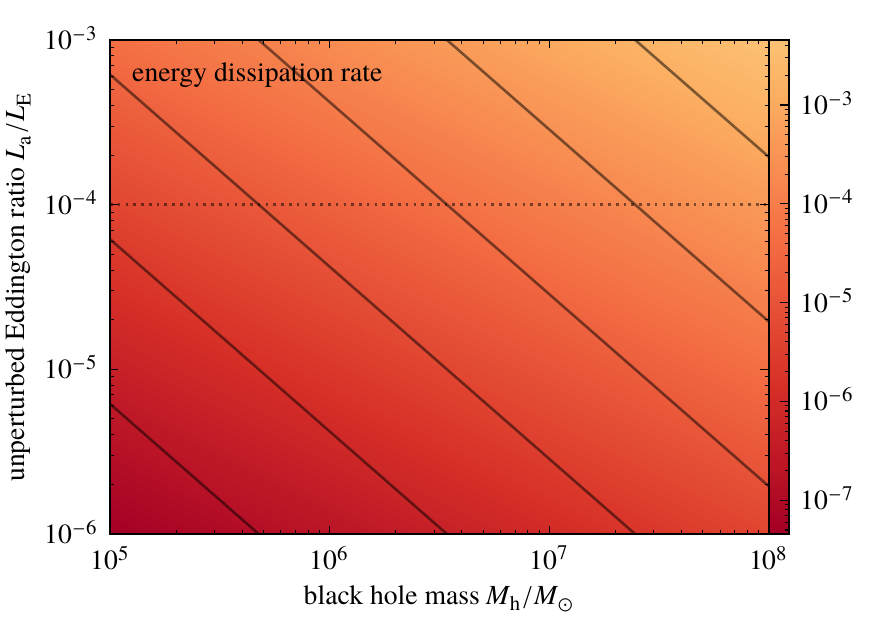}
\caption{Color contours of the initial energy dissipation rate $Q_0$ divided by
the Eddington luminosity $\LE$, as functions of the black hole mass $\Mh$ and
the Eddington ratio $\La/\LE$ of the unperturbed disk, with other parameters
fixed at their fiducial values (\cref{sec:summary,sec:light curves}). This
figure portrays the case in which the unperturbed inner disk is \iac{RIAF},
while \cref{fig:ss73} illustrates the case of a radiatively efficient disk.
Gray contours are drawn at constant levels of unperturbed $\mcr$, starting from
\num{e4} in the top-right and increasing by a factor of 10 for each contour
toward the bottom-left. The horizontal dotted line marks the value of $\La/\LE$
above which the \ac{RIAF} may become a radiatively efficient disk.}
\label{fig:riaf}
\end{figure}

The top panel of \cref{fig:ss73} displays $Q_0$ for a radiatively efficient
disk, and \cref{fig:riaf} does the same for \iac{RIAF}. We end \cref{fig:ss73}
at $\amc/\emc=\La/\LE=\num{e-4}$ and begin \cref{fig:riaf} at
$\amc/\emc=(\eta/\eta')(\La/\LE)=\num{e-2}$ in view of the uncertain
$\amc/\emc$ marking the transition between the two disk models. The ridge of
$Q_0$ follows from the fact that, in a radiatively efficient disk, $\Sigma_0$
reaches its maximum when disk pressure switches from gas- to
radiation-dominated. Because $Q_0\gtrsim\LE$ generally in a radiatively
efficient disk, there is enough dissipation to power Eddington-level
luminosity. By contrast, $v_R$ is much larger and $\Sigma_0$ much smaller in
\iac{RIAF} with the same $\amc/\emc$, so $Q_0\ll\LE$. A noticeable flare is
unlikely, so we drop the \ac{RIAF} from further consideration.

The middle panel of \cref{fig:ss73} plots the initial value of $\tinfl/\tcool$
for the radiatively efficient disk, with $\tinfl$ and $\tcool$ from
\cref{eq:inflow time,eq:cooling time} respectively. For most of the parameter
space plotted, $\tinfl/\tcool\lesssim1$, so $\Lc\sim\tfrac12\LE$ irrespective
of $\Mh$ and $\La/\LE$. With the gradual depletion of the inner disk,
$\tinfl/\tcool\propto1/\Sigma$ rises but $Q\propto\Sigma/\tinfl$ falls, and the
two changes offset each other to sustain a constant $\Lc$.

The luminosity plateau continues until $\tinfl/\tcool\sim1$. The plateau
duration is calculated by applying \cref{eq:plateau condition}, the condition
for a plateau. To connect $\tau$ in that equation to $\Sigma=\tau/\kappaT$ in
the simulations, we approximate $\Sigma(t)$ in the top-right panel of
\cref{fig:simulation} by
\begin{equation}\label{eq:mass decrease}
\Sigma/\Sigma_0\sim0.1\,(t/\tdec)^\gamma\quad\mtext{for $t\gtrsim\tdec$,}
\end{equation}
with $\tdec=\SI{\tdecfid}{\day}$ the start of the approximately power-law decay
and $\gamma=-2.5$ the power-law index. If $\tinfl/\tcool\gtrsim0.1$ initially,
this equation allows us to determine only an upper limit of
\SI{\lesssim\tdecfid}{\day} to the plateau duration. The plateau duration is
shown in the bottom panel of \cref{fig:ss73}, and explicit expressions can be
found in \cref{sec:plateau duration}. The plateau typically lasts tens of days.
After the plateau, the inner disk fades as $\Lc\sim Q\propto\Sigma$. For a
small part of the parameter space, $\tinfl/\tcool\gtrsim1$ even at the
beginning; these inner disks do not have a luminosity plateau.

The plateau duration is comparable to but typically smaller than $\tmb$.
Although the stream mass current decreases during the plateau, the disk mass
current would decrease by an even greater amount, according to \cref{eq:mass
decrease}. Therefore, the stream would become heavier and heavier relative to
the disk as the event progresses, so the assumption that energy dissipation in
the disk dominates that in the stream remains valid.

For a radiatively efficient unperturbed disk, \cref{fig:ss73} shows that the
maximum plateau duration of \SI{\sim23}{\day} is found at
$\Mh\sim\SI{4e6}{\solarmass}$ and $\La/\LE\sim\num{e-3}$. The disk giving rise
to the longest plateau is marginally radiatively efficient, and its pressure at
the stream impact radius is on the cusp between gas and radiation dominance;
such a disk has the greatest $\Sigma_0$ and thus the longest $\tcool$. The
numbers here are for the fiducial case of a Sun-like star, but the argument
holds for other main-sequence stars as well. As detailed in \cref{sec:plateau
duration} and illustrated in \cref{fig:plateau duration}, the maximum plateau
duration and the corresponding $\Mh$ increase with $\Ms$ as
$\mathrelp\propto(\Ms/\si{\solarmass})^{1-1/(12\gamma)}$ and
$\mathrelp\propto(\Ms/\si{\solarmass})^{7/8}$ respectively.

\begin{figure}
\includegraphics{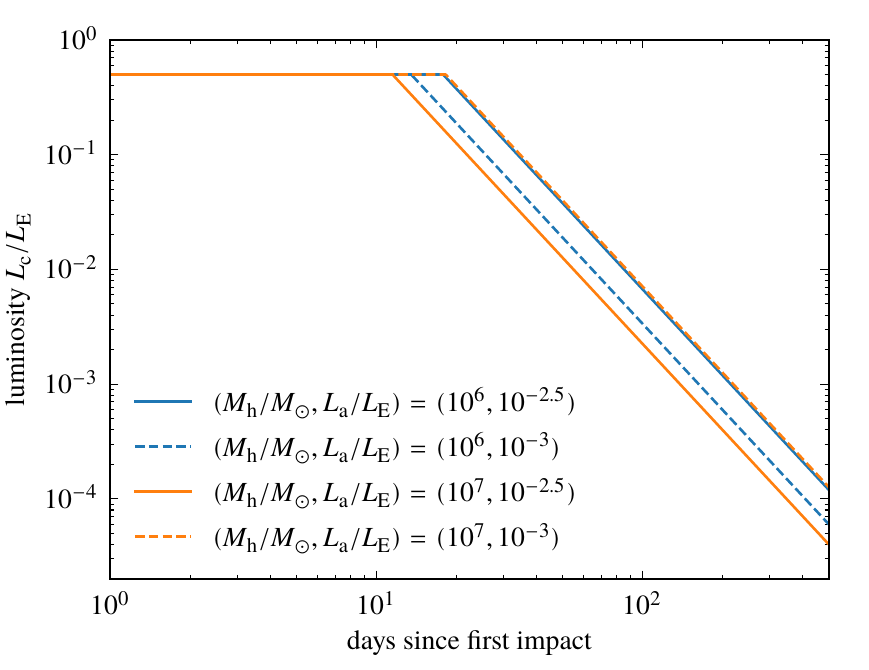}
\caption{Estimated bolometric light curves assuming that the unperturbed disk
is radiatively efficient. The level of the initial plateau is given by
\cref{eq:plateau}, and the plateau duration by the bottom panel of
\cref{fig:ss73}. That panel also shows, with crosses,
$(\Mh/\si{\solarmass},\La/\LE)$ of the depicted light curves. The plateau
duration depends weakly on $\Mh$ and $\La/\LE$, and the dependence changes sign
when disk pressure transitions from gas- to radiation-dominated
(\cref{sec:plateau duration}). The longest plateau duration for radiatively
efficient disks across all $\Mh$ and $\La/\LE$ is \SI{\sim23}{\day}
(\cref{fig:ss73}); the light curves here come close to that limit.}
\label{fig:light curve}
\end{figure}

\Cref{fig:light curve} displays the estimated bolometric light curves for four
radiatively efficient unperturbed disks. For all four, the luminosity plateau
lasts \SI{\sim20}{\day} to within a factor of \num{\sim1.5}. After the plateau,
$\Lc\sim Q\propto\Sigma\propto t^\gamma$. The power-law index of the falloff is
steeper than $-\tfrac53$, the power-law index of the mass return rate
\citep{1989IAUS..136..543P}, and the two are completely unrelated.

\section{Discussion}
\label{sec:discussion}

The Eddington-level luminosity plateau discussed here is qualitatively
different from the plateaus in other \ac{TDE} models. The classical picture
\citep[e.g.,][]{1988Natur.333..523R} describes \iac{TDE} around an isolated
black hole. The debris, returning to pericenter at super-Eddington rates for up
to several years \citep[e.g.,][]{1989ApJ...346L..13E}, quickly forms a circular
disk. The disk bolometric luminosity tracks the mass return rate, but instead
of going super-Eddington, it reaches a plateau
\citep[e.g.,][]{1997ApJ...489..573L, 2012ApJ...749...92K}. It should be noted
that the classical picture has been called into question by simulations
\citep{2015ApJ...804...85S} and their implications for observations
\citep{2015ApJ...806..164P, 2020arXiv200103234K}. In particular, the predicted
plateau is not seen in the light curves of most optical \acp{TDE}
\citep{2020arXiv200101409V}, and its existence is under debate.

The stream--disk interaction we discussed applies to \iac{TDE} around a black
hole with a pre-existing accretion disk. In contrast with the classical
picture, the plateau is produced when the debris slams into the disk, and the
principal parameter determining whether and for how long our plateau can be
observed is the accretion rate of the unperturbed disk. Moreover, the plateau
is shorter than in the classical picture, lasting only tens of days. One
respect in which our mechanism does resemble the classical picture, however, is
that the plateaus in both cases arise from radiation trapping, that is, a
situation in which inflow is faster than radiation can escape by diffusion.

\Citet{2019A&A...630A..98S} reported \iac{TDE} candidate in a quiescent galaxy
with a \SI{\sim e7}{\solarmass} black hole. The
\SIrange[range-phrase=-to-]{0.2}{2}{\kilo\electronvolt} flux held steady for
\SI{\sim90}{\day} at a \num{\sim e-3} Eddington level before diminishing as a
power law. If the galaxy were to harbor a weakly accreting black hole with no
recognizable \ac{AGN} signatures, then the disruption of a
\SI{\sim4}{\solarmass} main-sequence star could produce a plateau with the
observed duration. The dimness of the plateau may be explained by most of the
energy being radiated at other frequencies, or by dust extinction.

Our simulations did not account for magnetic fields. Magnetic field loops in
the outer disk can protrude into the depleted inner disk
\citep{2011ApJ...743..115N}, enhancing magnetic stresses to the degree that the
outer disk may replenish the inner disk on a timescale as fast as a few days.
Any resupply from the outer disk complements resupply from the stream, keeping
$\Sigma$ at a high level and prolonging the plateau.

Our calculations do not capture the gradual increase of the stream mass current
to its peak value in realistic \acp{TDE}. During the early parts of this rising
phase when the stream is lighter than the disk, our predictions here do not
apply, but shocks excited by the light stream can still deflect orbiting disk
material toward the black hole. Therefore, at later times when the stream is
heavier than the disk and our predictions do apply, the disk will have a
surface density smaller than its unperturbed value, suggesting that any
luminosity plateau that may appear would be shorter. However, if the net mass
loss from the inner disk during the rising phase is small due to resupply from
the stream or the outer disk, the plateau might be extended. Exactly how a
time-varying stream mass current affects the light curve is the subject of
future simulations.

Even though predictions about the spectrum of the inner disk are of great
astrophysical interest, we refrain from making such predictions because the
spectrum depends sensitively on the structure of the inner disk and its cooling
mechanisms, but we have little knowledge of either. The comparable magnitudes
of $\tinfl$ and $\tcool$ mean that gas and radiation energy densities in the
inner disk vary on similar timescales, and the vigorous energy dissipation in
the inner disk implies that radiation pressure may be instrumental in shaping
the inner disk. A careful treatment of the inner disk therefore requires
computationally expensive radiative \acp{MHD} simulations that can evolve gas
and radiation in unison. Radiative transfer must be performed in three
dimensions on account of the lack of symmetry: shocks are localized and heating
is uneven; the disk structure shown in \cref{fig:density} is highly irregular,
both along the midplane and in the vertical direction; and the stream arches
over and partly occludes the disk. It is also unclear to what degree free--free
and Compton processes bring the disk to thermal equilibrium. In light of all
these considerations, we content ourselves with an estimate of the light curve,
which already has a very distinctive shape.

\section{Conclusions}
\label{sec:conclusions}

The black hole of \iac{AGN} is surrounded by an accretion disk. The debris
stream of \iac{TDE} around such a black hole runs into the disk near
pericenter. The stream delivers a much stronger mass current to the impact
point than the disk does, so the dynamical evolution of the former is largely
unaffected by the latter. On the other hand, the heavy stream prevents the disk
from rotating freely. Shocks are formed where the disk smashes into the stream;
disk gas is deflected inward and drives shocks against gas closer in. As a
result, the kinetic energy of the disk interior to the impact point is
dissipated into internal energy at a rate high enough to power a bright flare.

If the unperturbed disk is radiatively efficient, that is, if its luminosity is
\num{\gtrsim e-3} Eddington, then the disk surface density $\Sigma$ is large
enough to make the initial energy dissipation rate $Q\propto\Sigma$
super-Eddington. Conversely, if the unperturbed disk is \iac{RIAF}, $\Sigma$ is
too small and $Q$ too low to beget any visible flare.

Because gas in the inner disk falls into the black hole in only a few orbits,
not all the energy dissipated issues forth as radiation. We estimate the
bolometric luminosity as $\Lc\sim Q\min(1,\tinfl/\tcool)$, where $\tinfl$ is
the time for a gas packet to flow inward from the impact point to the black
hole, and $\tcool$ is the cooling time of the inner disk. The luminosity at
early times is regulated to an Eddington-level plateau by the facts that
$Q\propto\Sigma/\tinfl$ and $\tcool\propto\Sigma$.

The luminosity plateau ends when $\tinfl/\tcool$ falls to \num{\sim1}, which
occurs after the disk has been depleted by shock-driven inflows. Resupply from
the stream and the outer disk could keep both $\Sigma$ and $\tcool$ large,
delaying the end of the plateau. If only stream resupply acts, as in our
simulations, then the plateau duration is largely determined by the initial
$\Sigma$, which is greatest for marginally radiatively efficient disks in which
gas and radiation contribute comparably to the pressure. The maximum plateau
duration is $\SI{\sim23}{\day}\times(\Ms/\si{\solarmass})$, produced by
\iac{TDE} around a $\SI{\sim4e6}{\solarmass}\times(\Ms/\si{\solarmass})^{7/8}$
black hole.

Following the end of the plateau, $\Lc\propto\Sigma$, which in our simulations
declines as a power law in time with index between $-3$ and $-2$ at earlier
times, and possibly a steeper index at later times.

Lastly, the simulations of \citet{2019ApJ...881..113C} show that part of the
stream drills through the disk and fans out on the other side into a dilute
plume, a fraction of which crashes back and collides inelastically with the
disk. The resulting luminosity enhancement, which could also be a fraction of
Eddington at early times, will be characterized in future work.

\ifapj\expandafter\acknowledgments\else\vskip\bigskipamount\noindent\fi
The authors thank Kojiro Kawana for the question that prompted this
investigation, and Tatsuya Matsumoto and Alexander Dittmann for fruitful
discussions. This work was partially supported by ERC advanced grant
\textquote{TReX} (CHC and TP) and NSF grant AST-1715032 (JHK).

\ifapj\software{Athena++ \citep{2008ApJS..178..137S}, Python, NumPy,
Matplotlib}\fi

\begin{appendices}

\section{Plateau duration}
\label{sec:plateau duration}

We present explicit expressions, applicable to all stars, for the plateau
duration when the unperturbed disk is radiatively efficient
\citep{1973A&A....24..337S}. Because $\rt$ depends on $\Ms$ and $\rs$, we
generalize $\tdec$ in \cref{eq:mass decrease} to
\begin{equation}
\tdec=\SI{\tdecfid}{\day}\times
  \biggl(\frac\Ms{\si{\solarmass}}\biggr)^{-1/2}
  \biggl(\frac\rs{\si{\solarradius}}\frac\rp\rt\biggr)^{3/2}.
\end{equation}
If the unperturbed disk is dominated by gas pressure and Thomson scattering
opacity at the stream impact radius, the plateau duration is
\begin{multline}\label{eq:gas plateau}
\SI{\sim14}{\day}\times
  \biggl(\frac\Ms{\si{\solarmass}}\biggr)^{-1/2-11/(30\gamma)}
  \biggl(\frac\rs{\si{\solarradius}}\frac\rp\rt\biggr)^
    {3/2+11/(10\gamma)}\times{} \\
\biggl[
  \biggl(\frac\Mh{\SI{e6}{\solarmass}}\biggr)^{-14/15}
  \biggl(\frac{\La/\LE}{\num{e-3}}\biggr)^{-3/5}
  \biggl(\frac\alpha{0.1}\biggr)^{4/5}
  \biggl(\frac\eta{0.1}\biggr)^{3/5}\biggr]^{1/\gamma},
\end{multline}
where $\gamma=-2.5$ from \cref{eq:mass decrease}. If the unperturbed disk is
dominated instead by radiation pressure and Thomson scattering opacity, the
plateau duration is
\begin{multline}\label{eq:radiation plateau}
\SI{\sim18}{\day}\times
  \biggl(\frac\Ms{\si{\solarmass}}\biggr)^{-1/2+1/(3\gamma)}
  \biggl(\frac\rs{\si{\solarradius}}\frac\rp\rt\biggr)^
    {3/2-1/\gamma}\times{} \\
\biggl[
  \biggl(\frac\Mh{\SI{e7}{\solarmass}}\biggr)^{2/3}
  \biggl(\frac{\La/\LE}{\num{e-3}}\biggr)
  \biggl(\frac\alpha{0.1}\biggr)
  \biggl(\frac\eta{0.1}\biggr)^{-1}\biggr]^{1/\gamma}.
\end{multline}
Because \cref{eq:mass decrease} is valid only for $t\gtrsim\tdec$, a plateau
duration shorter than that should be interpreted as an upper limit at $\tdec$.

\begin{figure}
\includegraphics{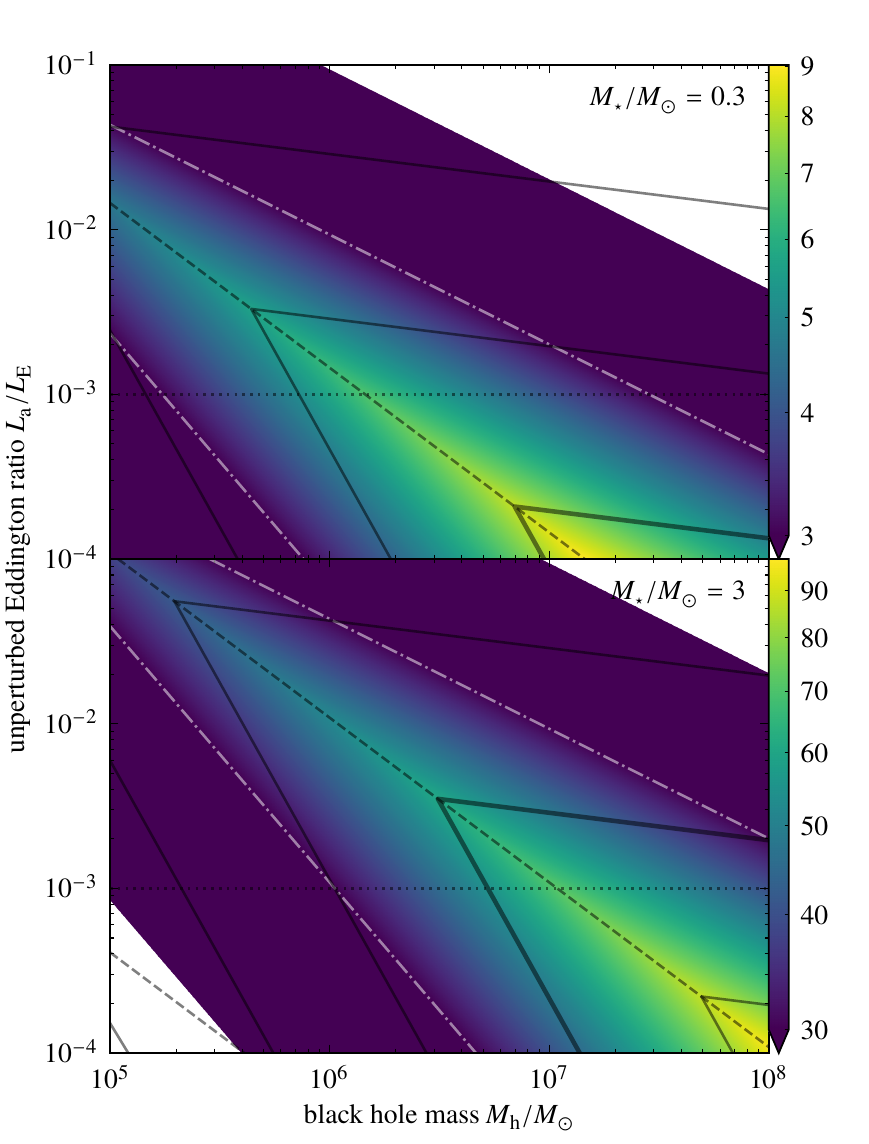}
\caption{Same as the bottom panel of \cref{fig:ss73}, but for different
main-sequence stars with masses as indicated in the top-right corner of each
panel. The plateau is longer for more massive stars. The gray chevrons are
contours of constant $\mcr$; they increase by factors of 10 toward the top
left, and the thick one is for $\mcr=1$. The contours slide upward with rising
$\Ms$. Our approximation that the stream has greater inertia than the disk
holds for $\mcr$ above a few tenths \citep{2019ApJ...881..113C}.}
\label{fig:plateau duration}
\end{figure}

\Cref{eq:gas plateau,eq:radiation plateau} are plotted in the bottom panel of
\cref{fig:ss73} for the fiducial case of a Sun-like star, and in
\cref{fig:plateau duration} for other main-sequence stars obeying
$\rs\propto\Ms$. The maximum plateau duration is
\begin{equation}
\SI{\sim23}{\day}\times
  \biggl(\frac\Ms{\si{\solarmass}}\biggr)^{-1/2+1/(24\gamma)}
  \biggl(\frac\rs{\si{\solarradius}}\frac\rp\rt\biggr)^{3/2-1/(8\gamma)},
\end{equation}
which is attained at
\begin{equation}
\Mh\sim\SI{4e6}{\solarmass}\times
  \biggl(\frac\Ms{\si{\solarmass}}\biggr)^{-7/16}
  \biggl(\frac\rs{\si{\solarradius}}\frac\rp\rt\biggr)^{21/16}.
\end{equation}

\end{appendices}

\ifapj\bibliography{tde}\fi
\ifarxiv\printbibliography\fi
\iflocal\printbibliography\fi

\end{document}